\documentclass{cup-ino}
\usepackage[utf8]{inputenc}
\usepackage{longtable}
\usepackage{blindtext,alltt}
\usepackage{float}
\RequirePackage{xpatch}
\title{AirCANS: CFD 2D Mesh Optimisation-based Airfoil Classification and Assessment using Neural Networks}

\author{Lushun Fan, Yuqin Xia, Jun Li, Karl Jenkins}

\begin{document}

\maketitle

\begin{abstract}
This study explores the possibilities of automating the loading, classification and assessment of Computational Fluid Dynamics (CFD) mesh data by Convolutional Neural Networks (CNNs). The research aim is finding a feasible way to quickly make classification and assessment on airfoil mesh data. For this purpose, this study designed a new framework named CFD-based airfoil Classification and Assessment Network (AirCANS) for CFD mesh data which including the data loader and improved the CNN structure to achieve our target. In our research, we found that CNNs are fully adaptable as well as understandable to CFD airfoil mesh data structures, which suggests that our hypothesis is successful and that neural networks can be used to have a greater positive impact on the CFD industry, such as it can be used to refine the mesh and accelerate the solution. This could allow CFD to spend much less time.
\end{abstract}

\section{Introduction}
Computational Fluid Dynamics (CFD) is widely used for simulating aerodynamic flows, climate dynamics, and industrial fluid processes \cite{runchal202050}. In CFD, resolving intricate flow structures in complex domains often requires mesh optimization to accurately capture small-scale flow features \cite{hoppe_view-dependent_1997, XuUCNN:Mesh}. However, the computational demands escalate substantially with mesh optimization, rendering fine-mesh CFD simulations highly sensitive to resource constraints in geometrically complex configurations. While traditional manual optimization methods remain time-intensive and technically challenging \cite{ALI2017431}, the development of efficient mesh optimization techniques that balance computational efficiency with accuracy has emerged as a critical frontier for advancing CFD capabilities \cite{doi:10.2514/6.2019-2945}. \\ 

The thickness of an airfoil plays a critical role in CFD design, as it corresponds to the flow field environment and thus influences  mesh optimization requirements. Therefore, using a neural network for the classification assessment of airfoil thickness is a technique of significant value, which can reduce research cycle time and costs by resolving the non-linear relationship between geometric features and aerodynamic layout \cite{Selig2003LowNotes}. The core significance lies in empowering real-time decision-making, supporting the assessment of thickness parameters to optimize airfoil adaptive performance in dynamic aerodynamic environments, and overcoming the limitations of engineers' experience-dependent judgment to tap into the laws that are difficult for humans to quickly discern, thus driving design innovation in airfoils\cite{Rayhan2024ComputationalGroove}. \\

Recently, deep learning has garnered considerable attention for its effectiveness in data-driven model development \cite{koeppe2021deep, Obiols-Sales2020CFDNet:Simulations, Calzolari2021DeepReview, Lim2024MeshLearning, WU2022AAdaptation}. Although deep learning has achieved significant advancements in areas such as image processing and pattern recognition, its potential for CFD mesh optimization \cite{Huang2021MachineDynamics} remains largely unexplored. A well-designed mesh should not only provide accurate results but also be computationally efficient. The optimization process involves adjusting the mesh geometry and resolution to minimize errors and decrease computation time. Traditionally, mesh optimization has been performed using deterministic methods such as gradient-based optimization or heuristic algorithms. However, these methods can be time-consuming and may not always produce optimal results. Deep learning offers a promising alternative for mesh optimization by leveraging deep neural networks \cite{Abucide-Armas2021ASimulations}. Deep learning algorithms have the ability to learn patterns from data and make predictions based on these learned relationships \cite{Raissi2017PhysicsEquations}. In the context of mesh optimization, deep learning algorithms can be trained on CFD simulation data to learn the relationship between mesh parameters and simulation accuracy \cite{Brunton2019MachineMechanics}. This knowledge can then be used to predict the optimal mesh for a given simulation. \\

While much of the existing research has focused on accelerating CFD simulations, this study emphasizes the efficient processing, classification, and analysis of mesh data files. Achieving this goal requires the development of neural network architectures specifically tailored to CFD mesh data. The primary aim of this research is to establish an innovative data-loading framework and design a dedicated neural network architecture for processing mesh data files in CFD applications. \\

The rest of this paper is structured as follows. The literature review section presents an overview of recent advancements in CFD techniques , especially mesh optimization using deep learning, and identifies current challenges . The methodology section describes in detail the generation of datasets, specifically , the generation of mesh data files, and outlines the processing methods used for CFD applications. It also addresses deep learning aspects, including the design of the data loader and modifications to the deep neural network, particularly adjustments made to the structure of the Convolutional Neural Network (CNN) to accommodate this specialized dataset. In the results section, we present and analyze the experimental results. In the final section, we discuss the limitations of the model, summarize the conclusions of the study, and outline future research directions and improvements.

\section{Literature Review}
Mesh generation is a very important step in CFD research. In the 2014 NASA report \cite{Slotnick2014CFDAerosciences}, geometry and mesh generation are described as one of the six important areas for future research. Therefore, many researchers focus on mesh optimization using deep learning.
Researchers often use PDEs in fluid dynamics \cite{Thuerey2018DeepFlows, Sirignano2018DGM:Equations}.\\

Mesh optimization \cite{HoppeMeshOptimization} is the process of changing a mesh, usually a triangle or tetrahedron mesh, to improve its quality or structure \cite{Lin2022AMeshes}. This process includes different tasks. Examples are smoothing the mesh \cite{Knupp2022GeometricParadigm}, adding or removing vertices, or moving vertices to better match features in the data. The main goal of mesh optimization is to create a mesh that accurately shows the original data \cite{Ahmed2021CFDReactor}. The optimized mesh should also have good numerical properties. These properties include good condition, small aspect ratio, fewer vertices, and good convergence. \\

In recent studies, some researchers have proposed new methods\cite{hanocka_meshcnn_2019, Qi2017PointNet:Segmentation}. These methods use Convolutional Neural Networks (CNNs) for the direct analysis of triangle meshes. Hancock et al. introduced  a CNN structure designed for the direct analysis of triangle meshes. Traditional 3D deep learning methods represent 3D shapes through multi-view projections.
These projections transform 3D data into regular 2D or 3D meshes. However, traditional methods often use too much memory and computing resources. They also do not preserve fine details well. Point cloud methods (such as PointNet and PointNet++) improve computational efficiency. However, these methods cannot clearly describe local structures. This issue makes it difficult to show complex shapes such as edges and surfaces.\\

With the development of CNN methods, one hybrid method \cite{barda_meshcnn_2021} with mesh collapse achieves high performance. MeshCNN works directly on the mesh edges. It uses convolution and pooling layers to capture the geometric information of the 3D mesh structure. The convolution step works on each edge and its four neighboring edges as one unit. This step helps MeshCNN better describe local geometric features and maintains invariance to rotation, translation, and scaling. In addition, the pooling layer simplifies meshes by folding edges. The network learns to choose which edges to remove during downsampling. This method simplifies meshes but keeps important shape features. This method performs better in experiments. On the SHREC shape classification dataset, MeshCNN achieves higher accuracy compared to traditional methods such as GWCNN \cite{Ezuz2017GWCNN:Analysis} and PointCNN \cite{Li2018PointCNN:Points}. In tasks such as COSEG and human body segmentation, the pooling step further improves model performance and makes results easier to understand by adaptively removing unimportant edges. This model remains stable and keeps good performance, especially under different samples of triangle meshes and changes in mesh structure. \\

However, researchers have not yet used neural networks to classify airfoil thickness and other parts of the flow field environment.
CNNs hold great promise for CFD applications. But researchers face big challenges when using CNNs to capture turbulent airflow patterns around airfoils. Current research heavily depends on structured mesh systems for experimental validation. However, structured mesh systems have limitations in adapting their meshes. Therefore, structured mesh systems are not suitable for complex geometries \cite{article}. Existing studies further limit their scope by using overly simplified homogeneous models. To address these limitations, this study use unstructured grid generation techniques to create experimental datasets. Then, we developed a new computing framework specifically designed to handle these mesh data.

The main innovations of this research are twofold. Firstly, we achieve an efficient classification of air film configurations by means of a novel collapsible pool architecture optimised specifically for airflow plane grid feature extraction, namely, AirCANS. Secondly, we have contributed to the field by publicly releasing a comprehensive CFD mesh dataset, thus facilitating comparative studies and advances in aerodynamic simulation methods. 



\section{Methodology}

\subsection{Overview}
\begin{figure}[!h]
    \centering
    \includegraphics[width=1\linewidth]{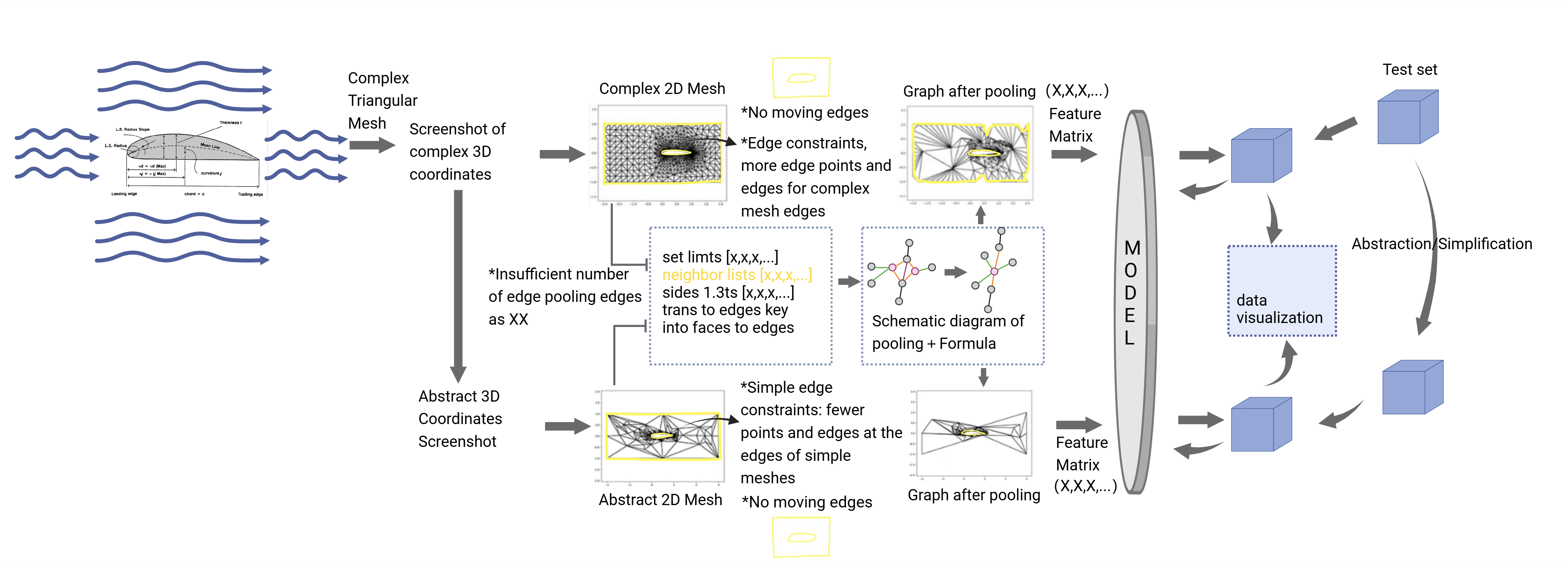}
    \caption{AirCANS framework: beginning with the input of the airfoil coordinates to generate simulation fluid meshes. The classification process concludes with construction, pooling, feature extraction and data visualization, encapsulating the transformation from geometric data to actionable insights.}
    \label{fig:flowchart}
\end{figure}

The pipeline, as shown in Figure~\ref{fig:flowchart}, changes raw airfoil shape coordinates into structured predictions using a geometric learning framework called AirCANS. We start from 2D surface outlines. Then, we create triangle meshes that divide the surrounding area into smaller parts. We preprocess these meshes to ensure structural regularity. For example, we make sure they meet edge count requirements for hierarchical operations. Once prepared, the model encodes meshes into edge-based features. Next, the model passes these features through convolution and pooling layers designed to capture local geometry and global topology. This process allows the model to abstract shape information step by step and predict mesh classification. We adapted AirCANS to a 2D plane to efficiently extract and identify airflow features. In addition, this study conducted comparative experiments to explore the need for mesh optimization in the next section. We evaluated classification efficiency between optimized and non-optimized meshes under complex mesh conditions. These experiments assess the model’s ability to handle complex airflow data in 2D scenarios. In particular, we studied how to identify and classify airflow during flying by observing airflow disturbances.\\

\begin{figure}
    \centering
    \includegraphics[width=0.5\linewidth]{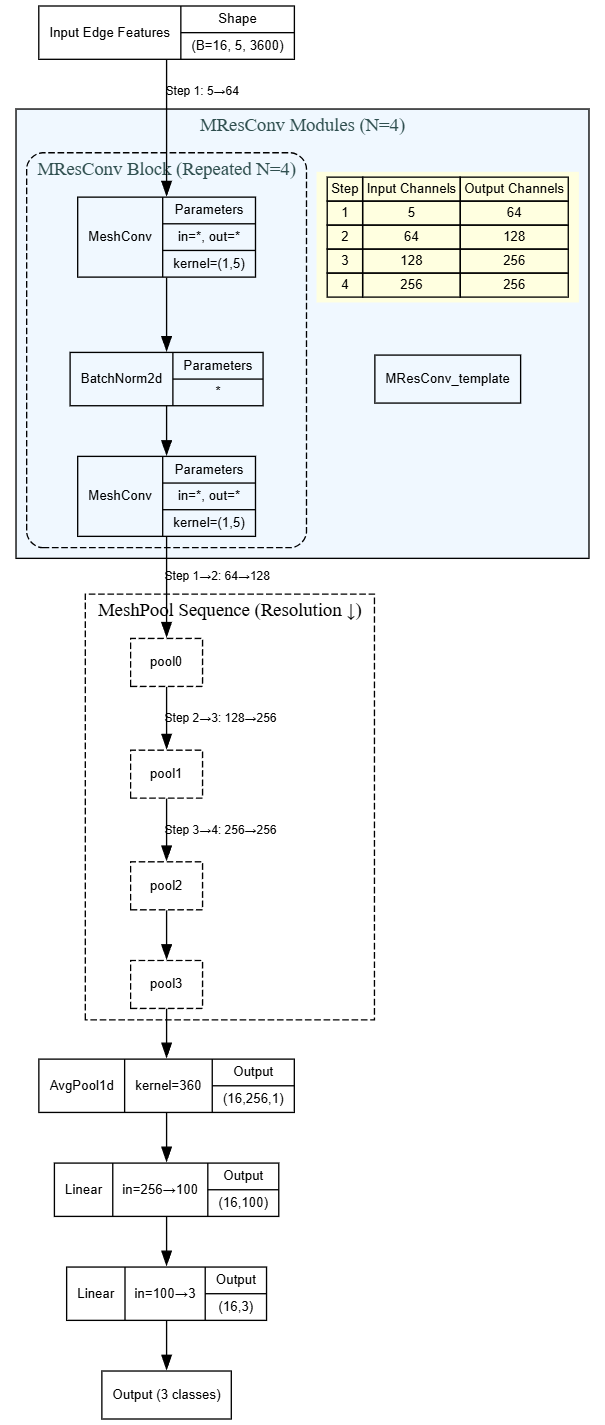}
    \caption{Architecture overview of the mesh-based classification network. 
The input edge features of shape $(B,5,E)$ are processed through a stack of MResConv modules composed of mesh convolutions, BatchNorm, and ReLU activations. Each stage increases the channel dimension while reducing the edge resolution via hierarchical MeshPool operations. A global average pooling layer aggregates edge features into a fixed-size representation, which is passed through two fully connected layers to produce the final class prediction.}
    \label{fig:Model and Data flow}
\end{figure}

The input to the network is an edge-based feature tensor of shape $\mathbb{R}^{B \times C_{\text{in}} \times E \times 1}$, where $B = 16$ is the default batch size, $E = 3600$ is the number of mesh edges, and $C_{\text{in}} = 5$ denotes the number of geometric features computed per edge. For each edge $e_i$, the feature vector is defined as
$
\left[\theta_i, \alpha_{i1}, \alpha_{i2}, r_{i1}, r_{i2} \right]
$
where $\alpha_{i1}$ and $\alpha_{i2}$ are the opposite angles within the neighboring triangles, and $r_{i1}, r_{i2}$ are the corresponding normalized edge length ratios within those triangles. $\theta_i$ is the dihedral angle between adjacent face normals, which is a meaningful descriptor in 3D surfaces. However, since our input meshes are strictly 2D, this component becomes constant and no longer contributes discriminative information. Despite this, we retain the full 5 dimensional feature format inherited from the original 3D network to maintain architectural consistency and enable possible fine-tuning or extension back to 3D data with minimal code modification. \\

The full input tensor $X \in \mathbb{R}^{B \times 5 \times E \times 1}$ is constructed by stacking all such feature vectors. The backbone network is composed of four hierarchical MResConv blocks. Each block contains two mesh convolution layers with kernel size $(1, 5)$, followed by BatchNorm2d and ReLU activation. The number of channels increases across layers as $C, 5 \rightarrow 64 \rightarrow 128 \rightarrow 256 \rightarrow 256$, while the number of edges is gradually reduced through mesh-based pooling as $E$, like $3600 \rightarrow 2100 \rightarrow 1800 \rightarrow 1500 \rightarrow 1200$ using MeshPool. \\

Each level concludes with GroupNorm to enhance training stability. After the final convolutional stage, a global average pooling layer aggregates features into a $[B, 256]$ tensor. This is passed through two fully connected layers to produce the final classification output. This design captures both local geometric structure and global shape semantics, making it effective for 3D mesh-based tasks.




\subsection{Data Generation}

\begin{figure}
    \centering
    \includegraphics[width=1\linewidth]{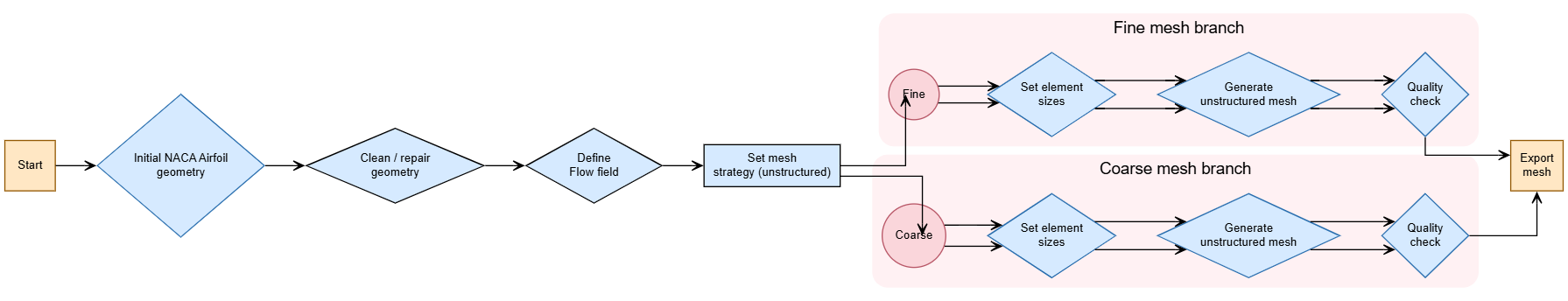}
    \caption{Mesh generation}
    \label{fig:mesh generation section}
\end{figure}

\begin{figure}[!h]
    \centering
    \includegraphics[width=0.8\linewidth]{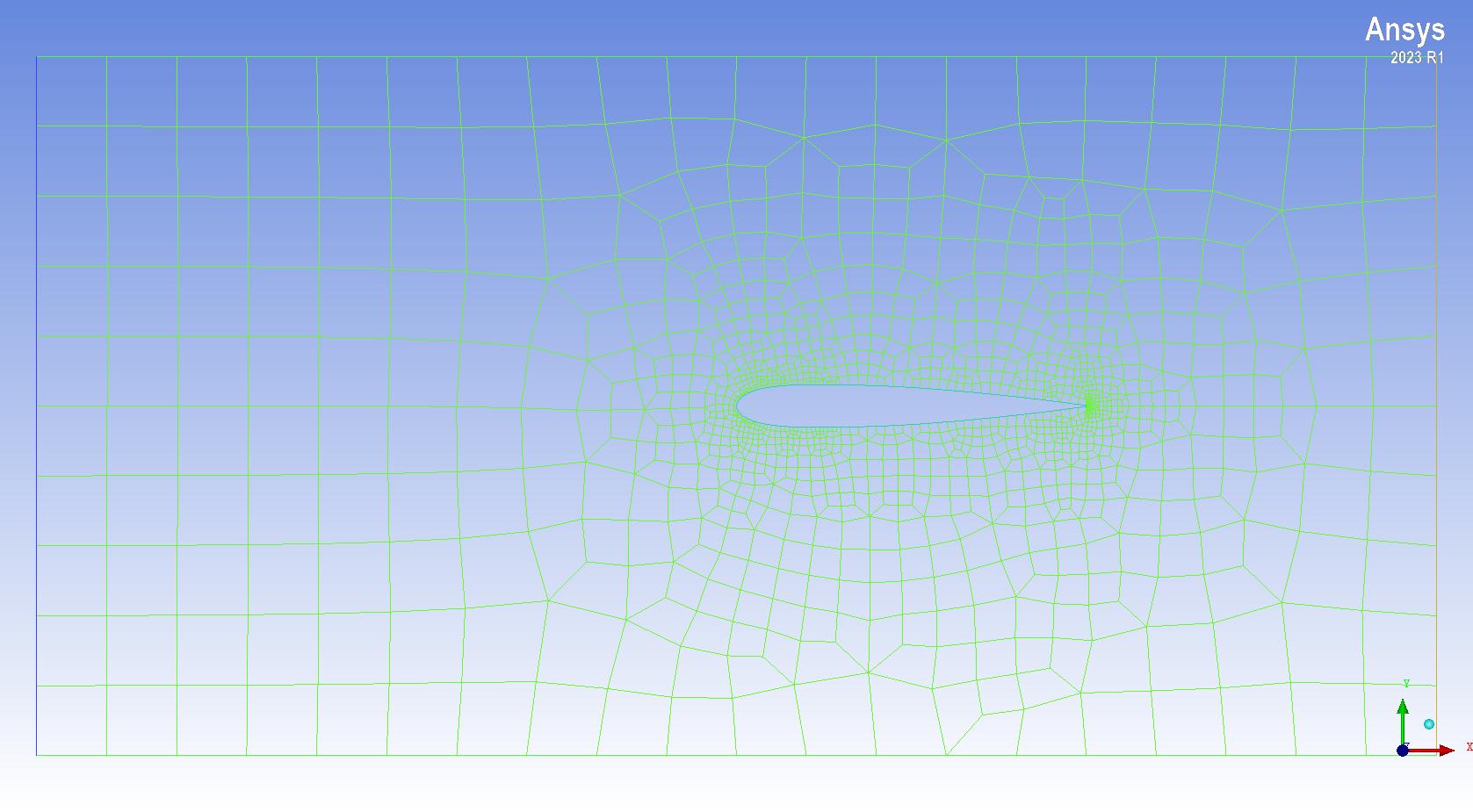}
    \caption{Airfoil Mesh}
    \label{fig:enter-label}
\end{figure}

Dataset generation is the first point of research innovation. It includes over 50 sets of Airfoil Mesh data files. This study generated a large number of randomly selected airfoil 2D unstructured mesh given like Figure \ref{fig:enter-label}. This is because structured meshes cannot be used in complex geometries, so we use unstructured meshes to cope with more complex geometrical possibilities. We divide all the airfoil mesh into 3 groups, such as airfoil thickness 1-10, 11-20 and 21-30. This is because, in aerodynamics, different airfoil thicknesses are usually applied in different flow field environments.The specific process is shown in the figure\ref{fig:mesh generation section}\\

\begin{figure}[!h]
    \centering
    \includegraphics[width=0.8\linewidth]{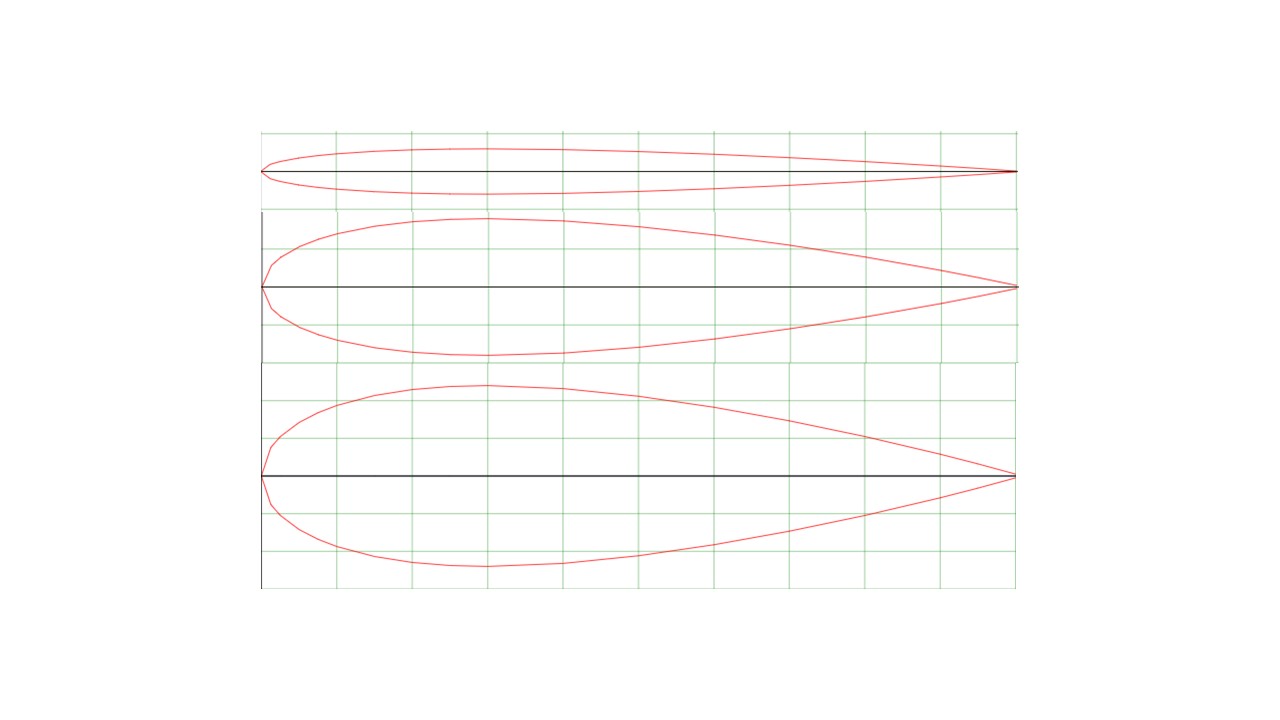}
    \caption{Multiple classifications of Airfoil samples. The vertical cross sections of the wing provide representative features unique to each airfoil type, capturing critical aerodynamic characteristics. This allows for effective classification based on 2D data without losing essential shape and flow details.}
    \label{fig:enter-label1}
\end{figure}

There are obvious differences in the performance of wings with different thicknesses: the thick wing type is suitable for low-speed flight with higher lift, good structural rigidity and large internal space, but the high-speed drag increases, making it difficult to realize supersonic flight; the thin wing type is suitable for high-speed flight with lower drag and excellent supersonic performance, but with lower lift coefficients, limited internal space, and higher requirements for structural design. The figure\ref{fig:enter-label1} shows the representatives of the three groups, from top to bottom are max thickness 6\%, 18\%, 24\% at 30\% chord. \\


The fundamental airfoil geometries utilized in this study were obtained from the public resource website Airfoil Tool. To facilitate mesh generation, unstructured meshes were employed due to their adaptability in handling complex geometries. This approach provides several advantages for advancing future research and experimental applications. However, the inherent characteristics of unstructured meshes make it challenging to precisely control the number of nodes. Consequently, it is essential to regulate the number of nodes across multiple dimensions by adjusting parameters such as the global mesh size and growth rate. \\



This study generated two distinct types of meshes: one fine and one coarse mesh for the airfoil, as Table \ref{tab:meshgenerator}. This evaluation criterion is the Jacobian Determinant in CFD, where fine mesh have mesh positives greater than 0 and mostly greater than 0.3, whereas coarse grids will have mesh distributions less than 0, which will make them unusable for simulations. This differentiation was designed to assess whether mesh resolution—fine versus coarse—exerts a significant influence on the performance and applicability of mesh files in neural network-based analyses. \\

\begin{table}[!ht]
\centering
\begin{tabular}{|l|l|l|l|l|}
\hline
\textbf{Mesh Density} & \textbf{Number of Layer} & \textbf{Initial Hieight} & \textbf{Height Ratio} & \textbf{Growth Law} \\ \hline
Airfoil Fine & 20              & 1               & 1.3          & explosion  \\ \hline
Coarse       & 10              & 1               & 1.1          & exposion   \\ \hline
\end{tabular}
\caption{This experiment uses Ansys ICEMCFD to generate parameter references of different density grids based on airfoil boundary coordinates}
\label{tab:meshgenerator}
\end{table}

Moreover, considering the primary components of AirCANNS are its convolutional and pooling layers, both of which operate based on kernel length, the number of edges $\textbf{E}$ generated need to be compatible with the kernel size $\textbf{K}$ while maintaining essential vertex connectivity of each vertex $\textbf{d(v)}$.

\[
E' = \left\lfloor \frac{E}{K} \right\rfloor \times K, \text{where } d(v) >= 2
\]


\subsection{Classification}
\begin{figure}[H]
    \centering
    \includegraphics[width=0.8\linewidth]{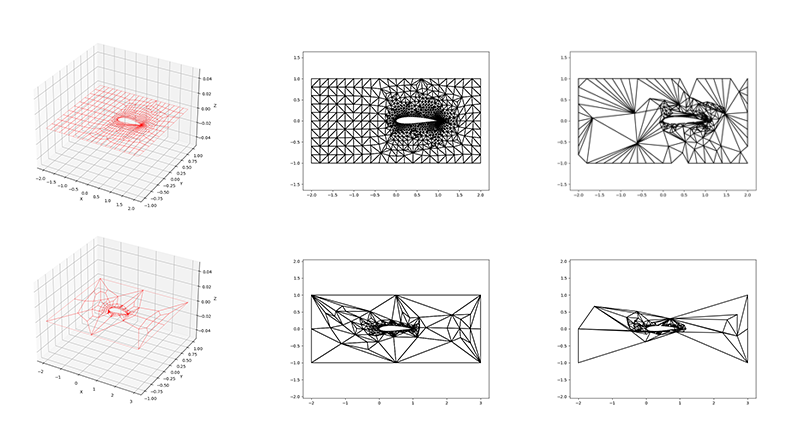}
    \caption{This figure illustrates the numerical simulation of airflow over an aircraft wing. From left to right, the top row shows the airflow discretized into a triangular mesh and subsequent collapse pooling. The bottom row presents a comparison experiment with an abstracted mesh grid in advance.}
    \label{fig:complexmesh}
\end{figure}

One common fundamental representation for airflow perturbations in fluid visualisation is the non-uniform polygonal mesh, with the mesh density showing the gas distribution. Through the mesh generating method (shown in Figure \ref{fig:complexmesh}), it is directly noted that the change of airflow around the wing, with an obvious blank area in the middle, corresponds to the fixed airfoil—serves position, which aligns with the boundary constraints (fixed boundary edges), while the other meshes indicate the direction and strength of the airflow. By analysing these mesh data, the model could identify and classify different shapes of airfoils in the flow field. This setup enables AirCANS to apply convolutional operations on the surrounding grid while maintaining the fixed-airfoil structure, facilitating effective feature extraction and classification. \\

\subsection{Neural network}
The neural network architecture, shown as figure \ref{fig:Model and Data flow}, performs convolution operations directly on the edges of meshes, leveraging the topology and geometric properties. The initial data for the mesh consists of a collection of vertex positions and triangular mesh vertex index triples, denoted as $(V, F)$. Subsequently, the geometric information is extracted to generate the edge set and geometric features, including edge neighbourhoods, set of sides, dihedral angles and ratios, used to assist the convolution pooling operation. Once the initial geometric features are extracted from the edges, the information in $V$ is no longer used in subsequent network computations. The network is concerned with how the features of the edges are transformed and transferred through edge-to-edge operations, rather than the actual shape of the mesh in space. \\

The most important layers of the neural network structure are the pooling layer and the non-pooling layer. The pooling and non-pooling operations employ an innovative mechanism called ‘edge folding’ based on Meshcnn, which is particularly suitable for dealing with irregular triangular mesh structures. Unlike traditional CNN pooling methods (e.g., simple region downsampling on an image), the pooling operation in MeshCNN simplifies the mesh by selectively collapsing the edges in the mesh, which reduces the complexity of the mesh while preserving key geometric features. \\

Specifically, the pooling process first determines the priority of folding by calculating the eigenvalues of each edge \cite{hanocka_meshcnn_2019}. Typically, edges with lower eigenvalues are preferentially collapsed during the pooling process, and these edges usually contribute less to the task. For example, in a classification task, edges in flat regions of the mesh tend to contribute less to shape recognition and can therefore be preferentially collapsed. When an edge is folded, its two neighboring triangles are merged into one, thus effectively preserving the geometry of the mesh while reducing the resolution. During the folding process, the features of the neighboring edges are also averaged to generate new edge features to retain more task-relevant information in the more simplified mesh. On the other hand, the non-pooling operation that shown as \ref{fig:Model and Data flow} is the inverse of the pooling process and aims to restore the original resolution of the mesh. In this process, storing historical information about edge folding during pooling (i.e., recording which edges were folded and how they were merged) enables the gradual recovery of pre-folding topology and edge features during subsequent non-pooling. The recovered feature data is then redistributed to each collapsed edge through averaging or interpolation operations, thus providing a high-resolution structure at the output that is consistent with the input. With the record of the mesh change history throughout the process, non-pooling can accurately reconstruct the original mesh topology, ensuring the accuracy of the output for tasks such as semantic segmentation. \\


The mathematical representation of the feature extraction to construct the initial matrix, convolution, and pooling operations is shown in following. For the Feature extraction, We will input the 2D datasets  \( I \in \mathbb{R}^{H \times W \times C} \) and extract the features like below: where $\nabla_x$ and $\nabla_y$ are the horizontal and vertical gradients. The gradients are then used to compute the gradient magnitude, which enhances local structural information. The filters also represent the boundary coordinates of the mesh along the x and y axes.
{Gradient Calculation:}
\begin{equation}
    \nabla_x I = I \ast S_x, \quad \nabla_y I = I \ast S_y
\end{equation}
where the Sobel filter is used in this research,
\begin{equation}
     S_x = \begin{bmatrix} -1 & 0 & 1 \\ -2 & 0 & 2 \\ -1 & 0 & 1 \end{bmatrix}, S_y = S_x^\top
\end{equation} 


{Gradient Magnitude at each pixel is computed as:}
\begin{equation}
    \left\| \nabla I \right\|_2 = \sqrt{(\nabla_x I)^2 + (\nabla_y I)^2}
    \label{3}
\end{equation}
{Initial Feature vector is composed of:}
\begin{equation}
    F^{(0)}_{i,j} = \left[ I_{i,j},\, \nabla_x I_{i,j},\, \nabla_y I_{i,j},\, \left\| \nabla I_{i,j} \right\|_2 \right]
\end{equation}
and later performs a convolution operation on it, for each pixel, $5\times5$ neighborhoods of $(i,j)$ to generate 7 sets of symmetric features : 
\begin{align}
    f_1 &= F^{(l)}_{i,j} \\
    f_2 &= \sum_{a=-1}^{1} \sum_{b=-1}^{1} F^{(l)}_{i+a,j+b} \quad &\text{(Local Sum)} \\
    f_3 &= \sum_{a=-2}^{2} \sum_{b=-2}^{2} F^{(l)}_{i+a,j+b} \quad &\text{(Extended Sum)} \\
    f_4 &= \left| \sum_{a=-1}^{1} F^{(l)}_{i+a,j} - \sum_{b=-1}^{1} F^{(l)}_{i,j+b} \right| \quad &\text{(Row-Column Difference)} \\
    f_5 &= \sqrt{\frac{1}{9} \sum_{a=-1}^{1} \sum_{b=-1}^{1} \left( F^{(l)}_{i+a,j+b} - \mu \right)^2}, \quad \mu = \frac{1}{9} \sum_{a,b} F^{(l)}_{i+a,j+b} \quad &\text{(Local Variance)} \\
    f_6 &= \max_{a,b \in \{-2,0,2\}} F^{(l)}_{i+a,j+b} \quad &\text{(Diagonal Max)} \\
    f_7 &= \min_{a,b \in \{-2,0,2\}} F^{(l)}_{i+a,j+b} \quad &\text{(Diagonal Min)}
\end{align}
and then by means of a convolution kernel \( W \in \mathbb{R}^{7C \times D \times 3 \times 3} \) calculate the output, 
Finally, the pooling operation includes factors as below,\\
Dynamic Importance Score:
\begin{equation}
    s_{i,j} = \alpha \cdot \left\| F_{i,j} \right\|_2 + \beta \cdot \left( -\sum_{k} p_k \log p_k \right)
\end{equation}
where $p_k = \frac{F_{i,j,k}}{\sum_{k} F_{i,j,k}}$, $\alpha$ and $\beta$ is a learnable parameter.\\
Merge \(2 \times 2\) regions with lowest \(s_{i,j}\) and Update adjacency indices for continuity.\\

Adaptive Weighted Average:
\begin{equation}
    F^{\text{pooled}}_{i,j} = \sum_{a,b=-1}^1 w_{a,b} \cdot F_{2i+a,2j+b}, \quad w_{a,b} = \frac{\exp(s_{2i+a,2j+b})}{\sum \exp(s_{2i+a,2j+b})}
\end{equation}
The convolutional layer is designed to operate on the edge features in the mesh structure, enhancing the feature representation by considering each edge jointly with the features of its four adjacent edges. This method employs symmetric functions for the neighbouring edge features to ensure the order invariance of the convolution operation, which is effective in adapting to the complexity and irregularity of the topology of the mesh data when dealing with it, thus improving the robustness of the model to geometrical changes. \\

With the topology of adjacent sides set, the pooling is executed through an edge-collapse mechanism, which is guided by the strength of edge features, quantified by their L2-norm as shown in Equation \ref{3}, involved merging an edge and its adjacent mesh elements into a single vertex. The process not only reduces the mesh complexity but also aggregates features from the edge and its geometric information, enhancing the representational efficiency of the network. However, it is noted that constraints of maintaining a closed topology after pooling inherently restrict the extent, particularly within the context of 2D representations, thereby limiting the degree to which the mesh can be compressed with its geometric and topological correctness. Consequently, this experiment considers the comparison groups to explore the exact influence which will be discussed in the next section.

\subsection{Optimization}

To investigate the impact of mesh density on model performance, we designed comparative experiments using both dense and sparse representations of the same airfoil CFD mesh. Our rationale was that the pooling layers in MeshCNN aim to reduce the number of edges to facilitate feature extraction. By providing a sparse mesh with fewer edges from the outset, we effectively performed a manual pooling in advance; the default parameters set in this project are shown in the above Table \ref{tab:meshgenerator}. This approach was intended to allow the model to focus its capacity more directly on extracting meaningful features, enhancing performance by reducing computational overhead associated with pooling operations during training. \\

By comparing the classification performance between these pre-simplified (sparse) meshes and the original complex (dense) meshes, we Our aim is to show the impact of mesh density properties in aerodynamics on data feature extraction in deep learning. This investigation helps determine whether manual mesh simplification can strike an optimal balance between classification accuracy and computational efficiency. The findings from these experiments are expected to provide insights into whether reducing mesh complexity prior to training enables the model to perform more effectively by concentrating on critical features pertinent to CFD analyses.



\section{Experiments}
We conducted multiple experiments to assess the efficacy of our design applied to computational fluid dynamics (CFD) mesh data. We developed the dataset with variations in mesh density to investigate the performance across different configurations of mesh complexity. 

\subsection{Experiments Setup}
AirCANS used Ansys ICEMCFD as generator software, Torch as the methodology environment, and customize MeshCNN \cite{hanocka_meshcnn_2019} to accommodate the specific characteristics of our dataset.. There are default parameters that refer to the work \cite{barda_meshcnn_2021}, like kernel size $K=3, pool_{size}=50$.
We designed experiments that systematically varied key parameters to understand their impact on model performance. Our primary objective was to determine how different training-to-testing dataset ratios, mesh representations, and learning rates affect the predictive accuracy of AirCANS framework when applied to CFD data. \\

\begin{figure}[!ht]
    \centering
    \includegraphics[width=0.8\linewidth]{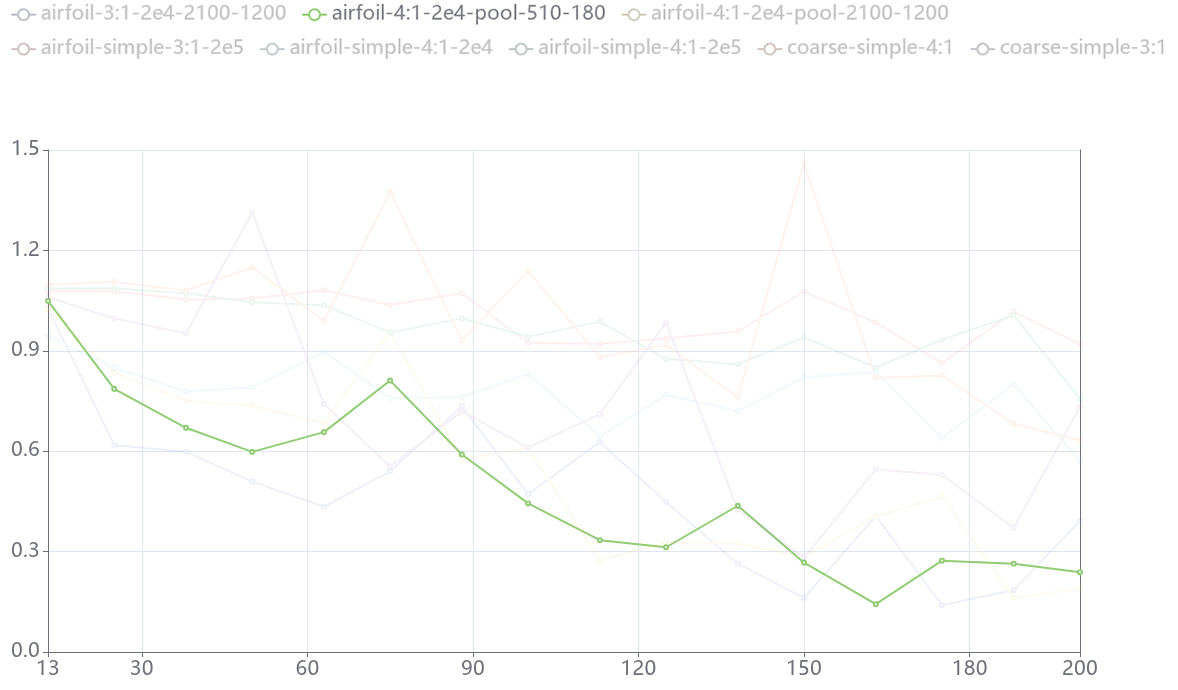}
    \caption{The figure compares training loss trajectories under different parameter configurations. Based on the convergence trends, we selected the pooling parameters $[510, 360, 210, 180]$ (derived from progressive downsampling) with a learning rate of $0.0002 (2×10^{-4})$, as this combination demonstrated the most stable and optimal loss reduction throughout the training process, outperforming alternative pooling configurations and learning rate settings.}
    \label{fig:lossgroup}
\end{figure}

Recognizing the critical role of training parameters and mesh density in machine learning models, we conducted comprehensive experiments to explore the influence of different parameters on model performance. We adjusted the training-to-testing ratios to examine how varying the amount of training data impacts the model's generalization and accuracy, as depicted in Figure \ref{fig:lossgroup}. Simultaneously, we assessed the effect of different mesh densities on feature extraction, using a consistent airfoil geometry but varying the level of detail from sparse to dense configurations. This included exploring critical hyperparameters that influence both the speed and quality of the learning process. By analyzing the results shown in the same figure, we were able to identify optimal conditions that enhance prediction accuracy and better understand the trade-offs between training data quantity, mesh density, and model performance.

\subsection{Results}


The results from the comparative experiments with varying data splits and network configurations are summarized in Table \ref{tab:optimized_comparison_exp}. We evaluated the impact of different train-test ratios and network complexity on the classification accuracy of airfoil datasets. Experiments using a more complex network structure with multiple convolutional layers [64,128,256,256] and multi-level pooling [510,360,210,180] achieved the best performance. Specifically, with a train-test ratio of 3:1, the highest accuracy reached during training was about 83\% (Highest ACC), while the model accuracy could be maintained at about 67\% at the later phase (Stable ACC). In contrast, simplified network architectures with fewer convolutional filters and pooling steps significantly degraded performance, achieving stable accuracies ranging from only about 33\% to 45\% and a maximum peak of 64\%. These findings demonstrate that even with a relatively small dataset, utilizing more sophisticated network architectures and an appropriate train-test ratio can considerably enhance the model's generalization and robustness. \\

\begin{table}[!h]
\centering
\caption{Comparison of Airfoil Experiments with Different Data Splits and Network Configurations}
\label{tab:optimized_comparison_exp}
\begin{tabular}{|l|c|c|c|c|c|c|}
\hline
\textbf{Ratio} & \textbf{lr} & \textbf{Network Filters (ncf)} & \textbf{Pooling} & \textbf{Stable Acc (\%)} & \textbf{Highest Acc (\%)} \\ \hline
3:1 & 2e-4 & [64,128,256,256] & [2100,1800,1500,1200] & 59.091 & 81.818 \\ \hline
4:1 & 2e-4 & [64,128,256,256] & [2100,1800,1500,1200] & 58.334 & \textbf{83.333} \\ \hline
4:1 & 2e-4 & [64,128,256,256] & [510,360,210,180] & 66.667 & \textbf{83.333} \\ \hline
4:1 & 2e-4 & [128] & [510] & 33.333 & 50.0 \\ \hline
4:1 & 2e-5 & [128] & [510] & 33.333 & 50.0 \\ \hline
3:1 & 2e-5 & [128] & [510] & 45.455 & 63.636 \\ \hline
\end{tabular}
\end{table}

Encouraged by these results, we further experimented with sparse mesh configurations to determine if reducing the initial number of edges would enhance the model efficiency and accuracy. In sparse data settings, presented in Table \ref{tab:sparse_comparison_exp}, based on the above result, the learning rate set as 2e-4 and train-test ratio as 4:1. Using a more complex network architecture consisting of multiple convolutional layers ([64,128,256,256]) and multi-level pooling resolutions ([450,420,390,360]), the stable accuracy after 200 epochs was 50\%, with the highest transient accuracy reaching 83\%. When using a simplified network configuration ([64,128] with fewer pooling resolutions [450,420]), the stable accuracy significantly decreased, although it once reached a peak accuracy of 83\%. \\

Comparing this with the previous non-sparse mesh experiments (Table \ref{tab:optimized_comparison_exp}), the results indicate that mesh sparsification maintained similar maximum peak accuracy (83\%), but showed lower stable performance (50\%) compared to the complex non-sparse mesh setup, which stabilized at 67\%. \\

\begin{table}[h]
\centering
\caption{Comparison of Sparse Airfoil Experiments with Different Data Splits and Network Configurations}
\label{tab:sparse_comparison_exp}
\begin{tabular}{|l|c|c|c|c|c|c|}
\hline
\textbf{Network Filters (ncf)} & \textbf{Pooling} & \textbf{Stable Acc (\%)} & \textbf{Highest Acc (\%)} \\ \hline
[64,128,256,256] & [450,420,390,360] & 50.0 & 83.333 \\ \hline
[64,128] & [450,420] & 33.333 & 83.333 \\ \hline
\end{tabular}
\end{table}

\begin{figure}[h!]
    \centering
    \begin{subfigure}[b]{0.48\textwidth}
        \centering
        \includegraphics[width=\textwidth]{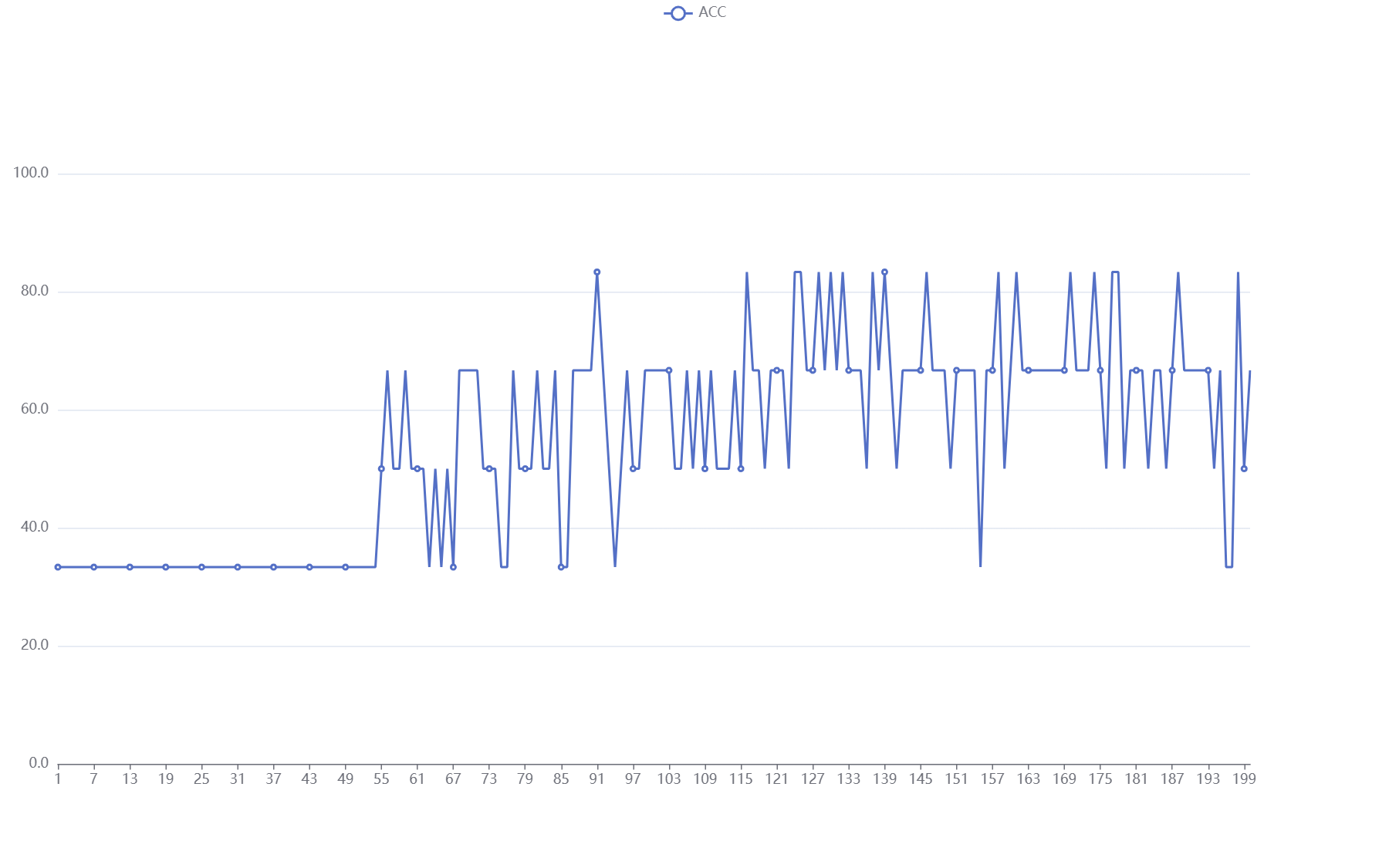}
        \caption{Dense grid results}
        \label{fig:dense_grid}
    \end{subfigure}
    \hfill
    \begin{subfigure}[b]{0.48\textwidth}
        \centering
        \includegraphics[width=\textwidth]{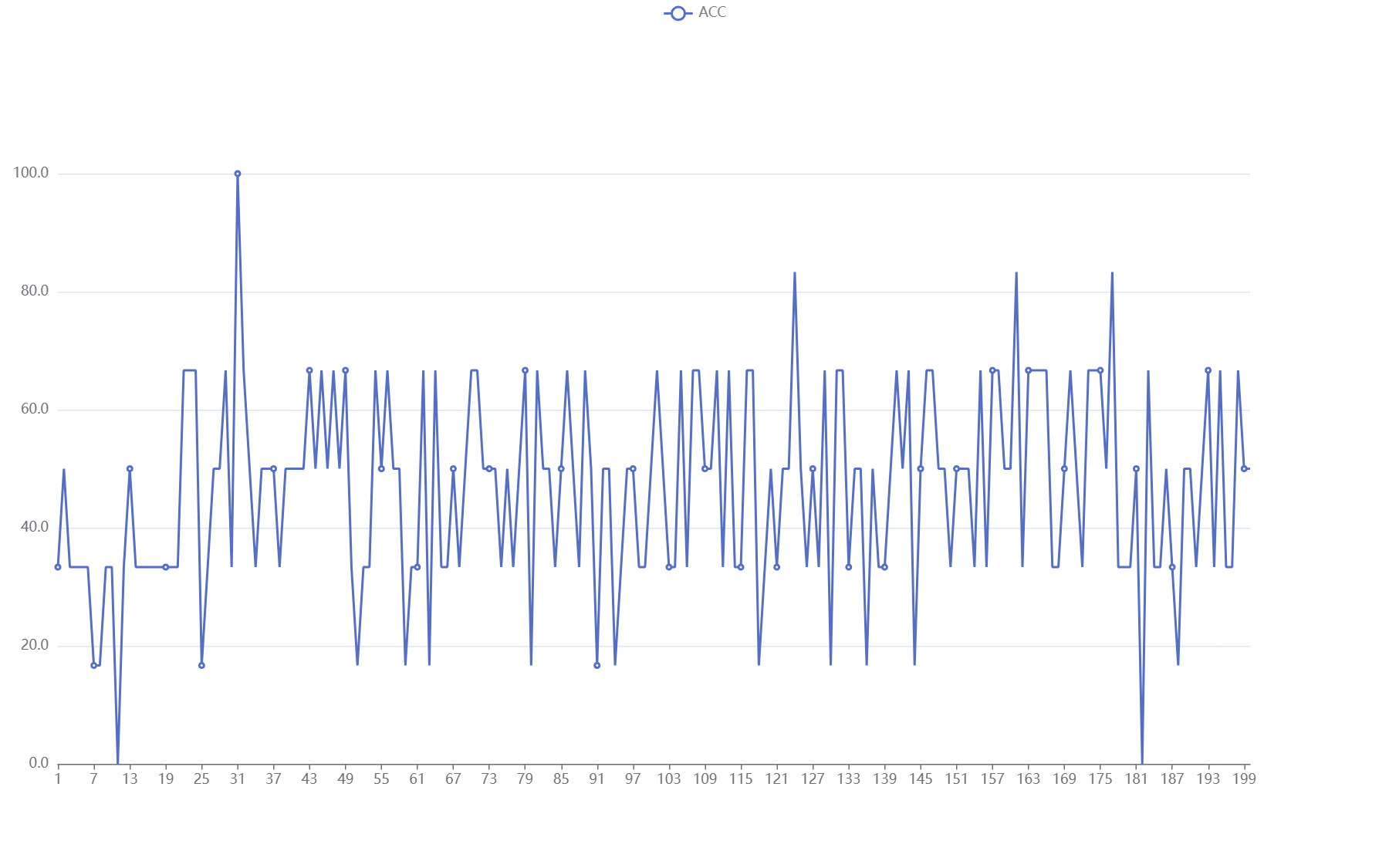}
        \caption{Sparse grid results}
        \label{fig:sparse_grid}
    \end{subfigure}
    \caption{Comparison of MeshCNN performance on different grid densities. Figure (a) shows the results on a dense grid, while Figure (b) illustrates the performance on a sparse grid. Note the high variability in the sparse grid results.}
    \label{fig:grid_results}
\end{figure}

Figures \ref{fig:dense_grid} and \ref{fig:sparse_grid} depict the accuracy outcomes for the same wing data, trained under varying conditions of mesh density during the 200 epoch train. Notably, despite our extensive parameter adjustments and balanced distribution of training and testing sets, the results remained consistent with those observed in previous trials. In the dense mesh configuration, with initial edges ranging from 2000 to 3000 and pooling parameters set to [2100, 1800, 1500, 1200] with a pool size of 50, the model demonstrates commendable stability, regularly achieving accuracy levels >80\%. Conversely, the sparse mesh setup, with fewer initial edges (500-800) and a reduced pool resolution [450, 420, 390, 360] with a pool size of 30, occasionally reaches peak accuracies of >99\% but exhibits significant fluctuations. This stark variability underscores the challenges and potential limitations inherent in training with highly abstracted mesh data. The development of the sparse model version, while capable of achieving exceptionally high accuracy scores, is notably less consistent than its dense counterpart, suggesting a trade-off between accuracy peaks and overall stability in model performance. \\

\section{Discussion}
In this research, developed a framework called AirCANS. This framework uses MeshCNN, a deep learning model specifically designed for classifying CFD meshes in two-dimensional datasets. Our model showed good performance in distinguishing among different mesh types with high accuracy. During training, the loss value consistently decreased from an initial value of 1.049 to a stable level around 0.238. This decrease shows that the model learned effectively. The accuracy of the model on test data remained high. These results offer a new method that improves analytical capabilities in this field. \\

The analysis of mesh representations demonstrated that mesh density has a significant impact on model performance. Models trained on denser meshes, which provide richer geometric information, outperformed those trained on sparser meshes. The increased detail in denser meshes facilitates better feature extraction, allowing the model to capture subtle flow characteristics around the airfoil. In contrast, the use of parse matrices—sparser mesh representations—led to a substantial decline in accuracy, even less than half. Moreover, adjusting the learning rate also did not yield the anticipated improvement. This suggests that sparse representations may inadequately capture essential features and disrupt the pooling operations, particularly near boundary conditions where accurate representation is critical.\\

There are several limitations to this study that may have affected the results. The complexity of data creation, due to the use of data we created ourselves, has prevented us from obtaining sufficient training data up to this point. And from the concept of mesh independence studies in this area of CFD, the number of mesh elements does not affect the simulation results after a certain number of mesh densities, but the reality is that the simulation results of the grids fluctuate in an interval because the Navier Stokes equations can only be solved in approximation at this point in time, so the results always fluctuate. Meanwhile, the diversity of the dataset could similarly affect the model's ability to learn the complex flow patterns that characterize CFD simulations. Insufficient training data can limit the learning capability, resulting in reduced prediction accuracy and generalization. The mesh representation is a key factor in model performance. On the other hand, given the boundary constraints of the model, the pooling layer requires soft-edge attribution; otherwise, the lack of geometric information may not adequately represent key features and may disrupt the pooling operation. This disruption is particularly severe near the boundary, as the abstracted mesh training pooling operation is limited this may also impose a limitation on model training.

\section{Conclusions and Future Work}
This study examines the application of MeshCNN on two-dimensional mesh datasets within the field of computational fluid dynamics, representing an integration of advanced neural network techniques in this area. The experiments focused on testing the MeshCNN model with different densities of mesh data. The results show variations in model performance with the initial setup only replacing the customised hypermeters based on our datasets, with dense meshes generally demonstrating more stable accuracy levels, whereas sparse meshes exhibited significant fluctuations in accuracy, including extreme highs and lows. \\

Future work should focus on optimizing these factors to further enhance model performance. Investigating alternative mesh representations that balance detail with computational efficiency may prove beneficial. Additionally, exploring adaptive learning rate strategies could improve convergence and robustness, potentially leading to more effective predictive tools in CFD applications. Such advancements would contribute to the development of innovative techniques in CFD analysis, leveraging the strengths of MeshCNN in processing and learning from complex geometric data.


\printbibliography

\section*{Key Words}
Convolutional Neural Network (CNN),

\end{document}